\documentclass[final,3p,times,twocolumn]{elsarticle}

 \usepackage{graphicx}
\usepackage{amssymb}
\usepackage{url}

\journal{Physics Letters B}

\begin{document}

\begin{frontmatter}

%% Title, authors and addresses

%% use the tnoteref command within \title for footnotes;
%% use the tnotetext command for the associated footnote;
%% use the fnref command within \author or \address for footnotes;
%% use the fntext command for the associated footnote;
%% use the corref command within \author for corresponding author footnotes;
%% use the cortext command for the associated footnote;
%% use the ead command for the email address,
%% and the form \ead[url] for the home page:
%%
%% \title{Title\tnoteref{label1}}
%% \tnotetext[label1]{}
%% \author{Name\corref{cor1}\fnref{label2}}
%% \ead{email address}
%% \ead[url]{home page}
%% \fntext[label2]{}
%% \cortext[cor1]{}
%% \address{Address\fnref{label3}}
%% \fntext[label3]{}

\title{Measurements of the strong-interaction widths of 
the kaonic $^3$He and $^4$He $2p$ levels}

%% use optional labels to link authors explicitly to addresses:
%% \author[label1,label2]{<author name>}
%% \address[label1]{<address>}
%% \address[label2]{<address>}

\author[]{(SIDDHARTA~collaboration)}
\author[lnf]{M.~Bazzi}
\author[victoria]{G.~Beer}
\author[milano]{L.~Bombelli}
\author[lnf,ifin]{A.M.~Bragadireanu}
\author[smi]{M.~Cargnelli}
%\author[lnf]{G. Corradi}
\author[lnf]{C.~Curceanu (Petrascu)}
\author[lnf]{A.~d'Uffizi}
\author[milano]{C.~Fiorini}
\author[milano]{T.~Frizzi}
\author[roma]{F.~Ghio}
%\author[roma]{B. Girolami}
\author[lnf]{C.~Guaraldo}
\author[ut]{R.S.~Hayano}
\author[lnf,ifin]{M.~Iliescu}
\author[smi]{T.~Ishiwatari\corref{cor1}}
\ead{tomoichi.ishiwatari@assoc.oeaw.ac.at}
\cortext[cor1]{T. Ishiwatari}
\author[riken]{M.~Iwasaki}
\author[smi,tum]{P.~Kienle}
\author[lnf]{P.~Levi Sandri}
\author[milano]{A.~Longoni}
%\author[lnf]{V.~Lucherini}
\author[smi]{J.~Marton}
\author[riken]{S.~Okada}
\author[lnf,ifin]{D.~Pietreanu}
\author[ifin]{T.~Ponta}
\author[lnf]{A.~Rizzo}
\author[lnf]{A.~Romero~Vidal}
\author[lnf]{E.~Sbardella}
\author[lnf]{A.~Scordo}
\author[ut]{H.~Shi}
\author[lnf,ifin]{D.L.~Sirghi}
\author[lnf,ifin]{F.~Sirghi}
\author[lnf]{H.~Tatsuno}
\author[ifin]{A.~Tudorache}
\author[ifin]{V.~Tudorache}
\author[tum]{O.~Vazquez~Doce}
\author[smi]{B.~W\"{u}nschek}
\author[smi]{E.~Widmann}
\author[smi]{J.~Zmeskal}

\address[lnf]{INFN, Laboratori Nazionali di Frascati, Frascati (Roma), Italy}
\address[victoria]{Dep. of Phys. and Astro., Univ. of Victoria, Victoria B.C., Canada}
\address[milano]{Politecnico di Milano, Sez. di Elettronica, Milano, Italy}
\address[ifin]{IFIN-HH, Magurele, Bucharest, Romania}
\address[smi]{Stefan-Meyer-Institut f\"{u}r subatomare Physik, Vienna, Austria}
\address[roma]{INFN Sez. di Roma I and Inst. Superiore di Sanita, Roma, Italy}
\address[ut]{Univ. of Tokyo, Tokyo, Japan}
\address[riken]{RIKEN, The Inst. of Phys. and Chem. Research, Saitama, Japan}
\address[tum]{Excellence Cluster Universe, Tech. Univ. M\"{u}nchen, Garching, Germany}

\begin{abstract}
The kaonic $^3$He and $^4$He X-rays emitted in the $3d \to 2p$ transitions were 
measured in the SIDDHARTA experiment. 
The widths of the kaonic $^3$He and $^4$He $2p$ states were determined to be
$
\Gamma_{2p}\left(^3{\rm He}\right) = 6 \pm 6 \mbox { (stat.)} \pm 7 \mbox{ (syst.) eV},
$
and 
$
\Gamma_{2p}\left(^4{\rm He}\right) = 14 \pm 8 \mbox { (stat.)} \pm 5 \mbox{ (syst.) eV},
$
respectively. 
Both results  are consistent with the theoretical predictions. The width
of kaonic $^4$He is much smaller than the value of $55 \pm 34$ eV determined by 
the experiments performed in the 70's and 80's, while the 
width of kaonic $^3$He was determined for the first time. 
\end{abstract}

\begin{keyword}
%% keywords here, in the form: keyword \sep keyword
 kaonic atoms \sep
 low-energy QCD \sep
 antikaon-nucleon physics \sep
 X-ray spectroscopy
%% PACS codes here, in the form: \PACS code \sep code
 \PACS
 36.10.k \sep
 13.75.Jz \sep
 32.30.Rj \sep
 29.40.Wk
%% MSC codes here, in the form: \MSC code \sep code
%% or \MSC[2008] code \sep code (2000 is the default)
\end{keyword}

\end{frontmatter}

%%
%% Start line numbering here if you want
%%
% \linenumbers

%% main text
\section{Introduction}
\label{sec:intro}

The X-ray measurements of the kaonic helium isotopes ($^3$He and $^4$He) play an important role
for understanding low-energy QCD in the strangeness sector. The measurements of kaonic $^4$He
X-rays performed in the 70's and 80's introduced a serious problem; {\it i.e.,} inconsistency between 
theory and experiment both in the shift and width of the kaonic $^4$He $2p$ state.

The $2p$ shift measured in the 70's and 80's \cite{Baird,Batty-NPA} was $-43 \pm 8$ eV on 
average, whereas theoretical calculations gave a shift below 1 eV based on
kaonic atom data with atomic numbers $Z \ge 3$ \cite{Batty-NPA,nuovo-ciemnto,Friedman-exa}. 
This discrepancy between theory and experiment was known as the ``kaonic helium puzzle''.

New results of the kaonic $^4$He $2p$ shift were recently obtained by the E570 \cite{e570} 
and SIDDHARTA \cite{sidd-khe4} experiments with a precision of a few eV. In addition, 
the shift of the kaonic $^3$He $2p$ state was determined by the SIDDHARTA experiment 
for the first time \cite{sidd-khe3}. The $2p$ level shifts both of kaonic $^3$He and $^4$He 
were found to be at most a few eV. Thus, the 
``kaonic helium puzzle'' of the shift was resolved.

Theoretically, a value of the widths of the $2p$ states of 
$\Gamma_{2p} =1 - 2$ eV was estimated both for kaonic $^3$He \cite{Friedman-exa} and 
$^4$He \cite{Baird,Batty-NPA,Friedman-exa}.

Experimentally, however, the width of the kaonic $^4$He $2p$ state was 
not well determined, leaving the situation unclear. The results for 
the $2p$ widths of $\Gamma_{2p} = 30 \pm 30$ eV \cite{khe2}, 
and $100 \pm 40$ eV \cite{Baird}, with an average of $55 \pm 34$ eV 
were reported by two groups, along with the following comment on 
their results \cite{Baird}. ``The shift measurements are seen 
to be in good agreement. The situation for the width values is much less satisfactory and 
the error bars of the two measured values do not overlap. The error on the quoted average 
has been taken from the external variance of the measured values.'' 

This discrepancy between measured width values and theory for 
the kaonic $^4$He is clarified by our measurement. 
We performed as well the first measurement of kaonic $^3$He 
with a similar precision.

\section{The SIDDHARTA experiment}
\label{sec:exp}

The kaonic helium X-rays were measured in the framework of the SIDDHARTA experiment 
performed at the DA$\Phi$NE electron-positron ($e^+e^-$) collider \cite{dafne}. 
The $\phi$(1020) resonance was produced at rest by the $e^+e^-$ collisions at 
the interaction point. The back-to-back correlated charged kaon pairs 
($K^+K^-$) from the $\phi$ decay were detected by the two scintillators mounted above and below 
the interaction region of the beam pipe 
(``kaon detector''). The coincidence signals of the $K^+K^-$ pairs 
were used for the timing selection of X-ray events. Cryogenic gas  
was used as a target. A cylindrical target cell made
of Kapton foils was filled with He gas.
In the top of the cell, thin Ti and Cu foils were installed 
for the energy calibration. 

X-rays were detected by large-area (1 cm$^2$) silicon drift detectors (SDDs). 
The SDDs, with a total area of 144 cm$^2$, were installed around the target cell.
The main background source at DA$\Phi$NE was charged particles scattered 
from the beams which were uncorrelated with the $K^+K^-$ coincidence. 
Thus, event selections using timing both of the $K^+K^-$ coincidence in 
the kaon detector and the X-ray hits on the SDDs 
suppressed background by about four orders of magnitude.

For details on the setup and experimental method we refer to \cite{kh-sidd}, 
where the result of the kaonic hydrogen measurement was reported.

The data of kaonic $^4$He measured in two periods (for one day each), 
and the data of kaonic $^3$He measured for four consecutive days in the 2009 data taking 
were used to extract the widths. Note that the data used for the shift 
determination in \cite{sidd-khe3} were reanalyzed to determine the strong-interaction width.

\section{Data analysis}
\label{sec:ana}

Calibration data were taken every several hours, by inserting  Ti and Cu foils and an X-ray tube below 
the setup to increase X-ray production rates. The energy scale of the X-ray data was 
calibrated for each SDD using the peak positions of the Ti K$\alpha$ (4.5 keV) and 
the Cu K$\alpha$ (8.0 keV) lines. The data were selected for further analysis in terms 
of the quality of stability, energy resolution and X-ray peak shapes.

In-beam data were categorized on the basis of timing with respect to the kaon trigger. 
The events with the kaon trigger are used for the analysis of kaonic atom X-ray signals, 
while those without the kaon trigger are used for the analysis of 
the energy scale and resolution of the SDDs.

\begin{figure}[htbp]
 \begin{center}
  \includegraphics*[width=1.0\linewidth]{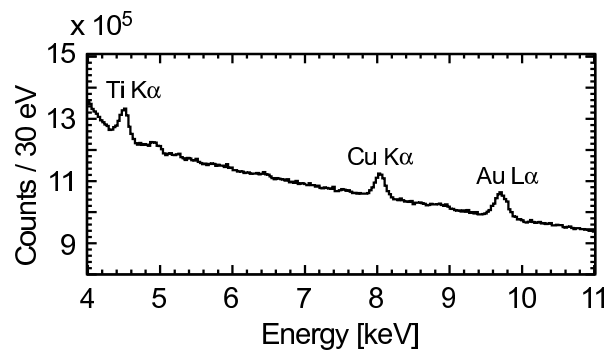}
  \caption{Energy spectrum filled with X-ray events  without the kaon trigger. 
The peak positions of the Ti K$\alpha$, Cu K$\alpha$, and Au L$\alpha$ are used for 
the determination of the energy scale.}
  \label{self}
 \end{center}
\end{figure}

An energy spectrum of the data taken with beam collisions is shown in Fig.~\ref{self}, 
where the events without the kaon trigger were selected. 
The background events originating from the beams are seen as a continuum. 
In addition, small peaks of the Ti K$\alpha$, Cu K$\alpha$, and Au L$\alpha$ 
lines are seen at 4.5, 8.0, and 9.6 keV. 
The Ti and Cu lines originated from the foils installed inside the target cell, while
the Au lines were from materials in the support structures of the SDDs.

The accuracy of the energy scale was examined using the X-ray energies of 
the Ti K$\alpha$, Cu K$\alpha$ and Au L$\alpha$ lines. 
The intensities of these X-ray lines in the kaonic helium data were 
not high enough to observe systematic effects on the energy calibration and 
energy resolution of the SDDs, so the data measured 
with the deuterium target, which were taken for about 20 days, 
were also analyzed using the same method.

The X-ray peaks were fitted with a Voigt function: $V=V(\sigma,\Gamma)=G(\sigma) \otimes L(\Gamma)$, 
which is a convolution of Gaussian $G(\sigma)$ and Lorentzian $L(\Gamma)$ functions.
The values of $\sigma$ and $\Gamma$ represent the width of Gaussian $G$ and Lorentzian $L$,
respectively. The detector response function was assumed to be Gaussian, and the natural linewidth
was represented by the Lorentzian function. In the fit, the Voigt function was calculated 
using the algorithm given in \cite{voigt}.

%Because the energy separation of the fine structure between the $\alpha_1$ and $\alpha_2$
%lines is much smaller than the detector resolution, the intensity ratio, energy difference, 
The intensity ratio, energy difference, 
%and  natural linewidths were fixed in the fit, where the values given in \cite{ti-ene,cu-ene,au-ene} were used.
and  natural linewidths of the fine structure between the $\alpha_1$ and $\alpha_2$ lines
were fixed in the fit, where the values given in \cite{ti-ene,cu-ene,au-ene} were used.

\begin{figure}[ht]
 \begin{center}
  \includegraphics*[width=0.45\textwidth]{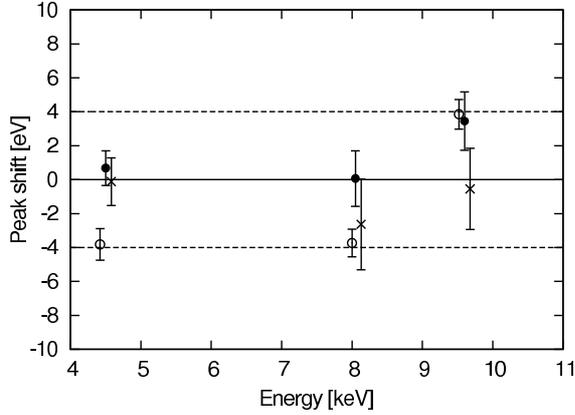}
  \caption{
The vertical axis gives the difference between the fit values and the reference data
at the peak positions of the Ti K$\alpha$ (4.5 keV), Cu K$\alpha$ (8.0 keV), and Au L$\alpha$ (9.6 keV) lines.
The peak positions of the X-ray lines are plotted separately
for different target materials: open circle (deuterium), filled circle ($^3$He),
and cross ($^4$He). The variation of the data points is within $\pm 4$ eV.
}
  \label{fig2}
 \end{center}
\end{figure}

Figure \ref{fig2} shows the peak positions of the Ti K$\alpha$ (4.5 keV), Cu K$\alpha$ (8.0 keV), 
and Au L$\alpha$ (9.6 keV) lines, where they are plotted separately for different target materials: open circle (deuterium), filled circle ($^3$He),
and cross ($^4$He).  The vertical axis gives the difference between the fit values and the reference data.
The variation of $\pm 4$ eV seen in the figure corresponds to the uncertainty of about $\pm 0.2$ channels 
in the analogue-to-digital converters (ADCs) used in the measurements. The variation could be 
related to the non-linearity of ADCs. Thus, the uncertainty of $\pm 4$ eV is taken as a systematic error 
in the energy determination. 

\begin{figure}[ht]
 \begin{center}
  \includegraphics*[width=0.45\textwidth]{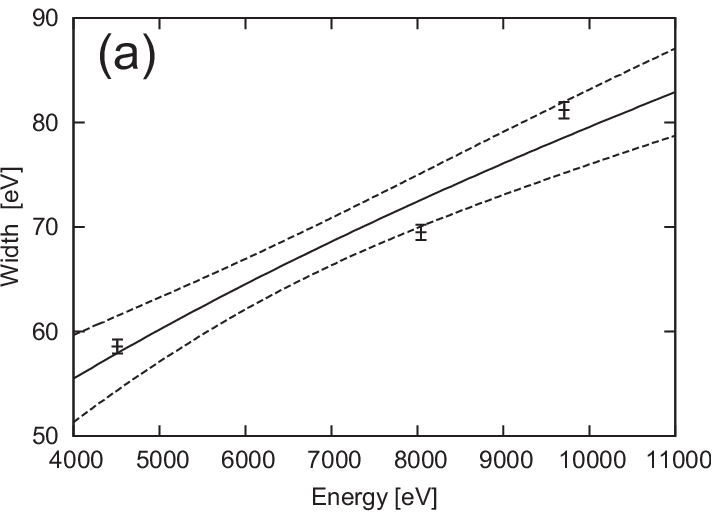}
  \includegraphics*[width=0.45\textwidth]{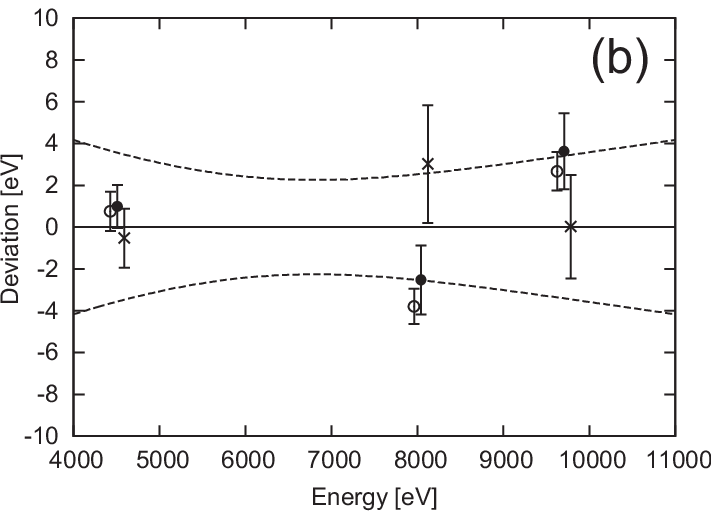}
  \caption{ (a): The Gaussian width $\sigma$ in the Voigt function at 
the Ti K$\alpha$ (4.5 keV), Cu K$\alpha$ (8.0 keV), and Au L$\alpha$ (9.6 keV) lines are plotted,
where the data of all the target materials were used. 
The fit curve using the function (\ref{width-fit}) is shown as a solid line, and 
the root-mean-square error of the fit as dotted lines.
(b): The deviations from the fit line plotted for each target material separately:
open circle (deuterium), filled circle ($^3$He), and cross ($^4$He). The curves 
show the uncertainty of the determination of the $\sigma$ values in (a).
}
  \label{width}
 \end{center}
\end{figure}

The energy dependency of the energy resolution of the SDDs was evaluated from the peak widths. 
Figure \ref{width}(a) shows the fit values of the Gaussian width $\sigma$ in the Voigt function 
against the X-ray energy $E$, where the data of all the target materials were used. 
The peak positions of the Ti K$\alpha$ (4.5 keV), Cu K$\alpha$ (8.0 keV), and Au L$\alpha$ (9.6 keV) lines are plotted.
The value of $\sigma(E)$ at the X-ray energy  $E$ can be expressed as:
\begin{equation}
\sigma(E) = \sqrt{a + b E},
\label{width-fit}
\end{equation}
using free parameters $a$ and $b$.
In Fig. \ref{width}(a), the fit curve using the function (\ref{width-fit}) 
is shown as a solid line and, as well, the root-mean-square error of 
the fit as  dotted lines.

The deviations from the fit line are plotted for each target material separately in Fig.~\ref{width}(b):
open circle (deuterium), filled circle ($^3$He), and cross ($^4$He). The curves show the error of 
the determination of the $\sigma$ values. Since all the positions are located within the error curves, 
the error is taken as the accuracy of the determination of the detector resolution for the fit of 
the kaonic helium X-rays. The energy resolutions ($\sigma$) at the X-ray energy of the kaonic helium 
$3d \to 2p$ transitions were determined to be : $\sigma = (65.4 \pm 2.3) \mbox{ eV}$ for kaonic $^3$He, and
$\sigma = (66.4 \pm 2.3) \mbox{ eV}$ for kaonic $^4$He.

In addition to the X-ray energy data, the time difference 
between the kaon coincidence and X-rays was measured, as well as 
the kaon time-of-flight of the kaon detector. The X-ray events 
were selected using this timing information, to obtain a good 
signal-to-background ratio in the energy spectra of kaonic atom X-rays 
without reducing their statistics \cite{sidd-khe3,kh-sidd}.

\begin{figure}[htb]
  \begin{center}
    \includegraphics[width=1.0\linewidth]{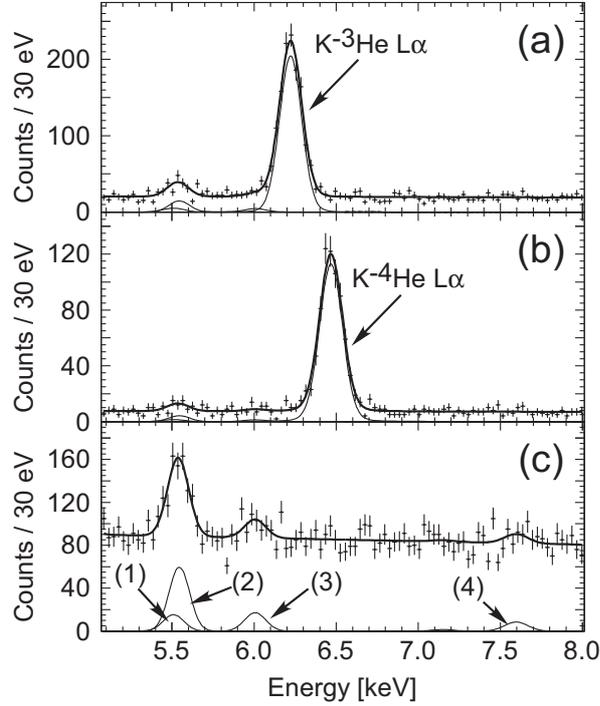}
    \caption{X-ray energy spectra of (a) kaonic $^3$He, (b) kaonic $^4$He, and
(c) kaonic deuterium. The thin lines show the peak fit functions after the background 
subtraction. The positions of the kaonic $^3$He and $^4$He $3d \to 2p$ transitions
are shown. In Fig. (c), (1): kaonic carbon $6 \to 5$ transition, 
(2): kaonic carbon $8 \to 6$ transition, (3): kaonic oxygen $7 \to 6$ transition, 
and (4): kaonic nitrogen $6 \to 5$ transition.}
    \label{fig:spec}
  \end{center}
\end{figure}

The energy spectra of the kaonic $^3$He and $^4$He X-rays are shown in Figs.~\ref{fig:spec}(a) and (b),
where the thin lines show the peak fit functions after the background subtraction.
The peaks at 6.2 keV and 6.4 keV are the kaonic $^3$He and $^4$He 
$3d \to 2p$ transitions, respectively.
Figure ~\ref{fig:spec}(c) shows the X-ray energy spectrum using the deuterium target, where
the signals from the kaonic deuterium X-rays are not visible. The upper limit of the observation
of kaonic deuterium will be reported elsewhere \cite{kd-paper}.

In addition to kaonic helium, several small peaks were observed in all 
the spectra, which originated from kaonic atom X-rays produced in the target window material made of Kapton Polyimide 
(${\rm C}_{22}{\rm H}_{10}{\rm N}_{2}{\rm O}_{5}$), since some kaons 
are stopped there.
The X-ray peaks at 5.5, 6.0, and 7.6 keV are the kaonic carbon ($K^-$C) $6 \to 5$, oxygen ($K^-$O) 
$7 \to 6$, and nitrogen ($K^-$N) $6 \to 5$ transitions, respectively.

In these transitions, the shift and broadening due to the strong-interaction are negligibly small \cite{optical}. 
Thus, their peak positions can be calculated using the QED effect only, as shown in Table \ref{tab:calc}. 
The energy shift caused by the vacuum polarization effect was obtained using the formula given in \cite{ue}, 
where the first order of the Uehling potential was taken into account. For the $\Delta n = 2$ transition
($K^-$C $8\to6$), the formula given in \cite{ue2} was used. The contribution from higher order 
corrections is estimated to be within 0.2 eV.

\begin{table}[htbp]
\caption{Calculated energy of kaonic atom X-rays.}
\label{tab:calc}
\begin{center}
\begin{tabular}{ccc} \hline \hline
Target & Transition & Energy (eV) \\ \hline
C & $8 \to 6$  & 5510 \\
C & $6 \to 5$  & 5545 \\
O & $7 \to 6$  & 6007 \\
$^3$He & $3 \to 2$ & 6225 \\
$^4$He & $3 \to 2$ & 6463 \\
Al& $9 \to 8$  & 7151 \\
N& $6 \to 5$  & 7595 \\
\hline
\hline
\end{tabular}
\end{center}
\end{table}

On the other hand, a small shift and a small width could be expected for the kaonic 
helium $2p$ states, due to the strong-interaction between the kaon and helium, whereas 
they are negligible in the $3d$ states. The energy shift and broadening of the $3d \to 2p$ transition 
can be obtained from the peak fit using a Voigt function: $V=V(\sigma,\Gamma)$, 
where $\Gamma$ represents the strong-interaction $2p$ width. Due to a strong parameter correlation 
between the values of $\sigma$ and $\Gamma$ in the fit, the value of $\sigma$ was 
fixed using the value obtained  from equation (\ref{width-fit}).

The determined width of kaonic $^3$He $2p$ state is:
\begin{equation}
\Gamma_{2p}\left(^3{\rm He}\right) = 6 \pm 6 \mbox { (stat.)} \pm 7 \mbox{ (syst.) eV},
\label{he3-gamma}
\end{equation}
and  the width of kaonic $^4$He $2p$ state is
\begin{equation}
\Gamma_{2p}\left(^4{\rm He}\right) = 14 \pm 8 \mbox { (stat.)} \pm 5 \mbox{ (syst.) eV}.
\label{he4-gamma}
\end{equation}
Here, the systematic error was evaluated from the uncertainty of the resolution $\sigma$.
In addition, since the kaonic $^3$He X-ray peak partially overlaps the $K^-$O $7 \to 6$ transition,
the systematic error of the width of kaonic $^3$He includes the uncertainty of its intensity.
Other contributions are small compared to the assigned systematic errors. 

\if0
The determined widths are very small compared to the detector resolution.
A slight deviation from the Gaussian shape in the detector response may cause a small 
positive central value in the width determination. The accuracy of the central value 
was found to be low for the case of a small width.
\fi

Note that the shifts are the same as the values reported in \cite{sidd-khe3}: 
$-2 \pm 2 \mbox { (stat.)} \pm 4 \mbox{ (syst.) eV}$ for kaonic $^3$He, and
$+5 \pm 3 \mbox { (stat.)} \pm 4 \mbox{ (syst.) eV}$ for kaonic $^4$He.

\section{Conclusion and discussion}
\label{concl}

In conclusion, the strong-interaction widths both of the kaonic $^3$He and $^4$He $2p$ states
were measured by the SIDDHARTA experiment, where 
kaonic $^3$He was measured for the first time.
The width of kaonic $^4$He was found to be much smaller than 
the value of $55 \pm 34$ eV determined in 
the experiments performed in the 70's and 80's \cite{Baird,Batty-NPA}.
 The strong-interaction $2p$ level widths both of kaonic $^3$He and $^4$He are in 
good agreement with the theoretical estimated values of 1-2 eV \cite{Baird,Batty-NPA,Friedman-exa}.
No abnormally large widths were found either in kaonic $^3$He or $^4$He.

\begin{figure}[htb]
  \begin{center}
    \includegraphics[width=1.0\linewidth]{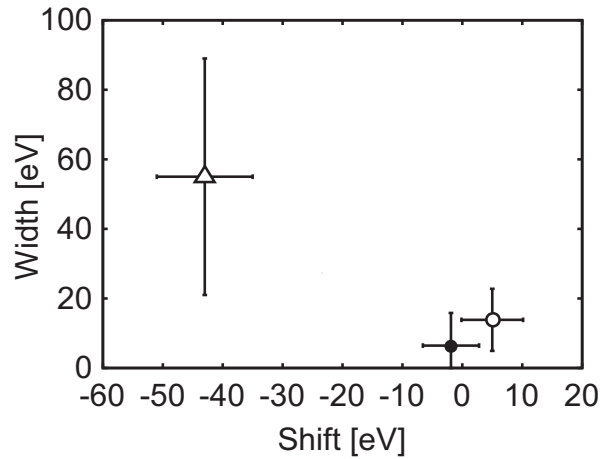}
    \caption{Comparison of experimental results. 
Open circle: K-$^4$He $2p$ state; filled circle: K-$^3$He $2p$ state.
Both are determined by the SIDDHARTA experiment. 
The average value of the K-$^4$He experiments performed in the 70's and 80's 
is plotted with the open triangle.}
    \label{results}
  \end{center}
\end{figure}

Combined with the results of the shift values determined in \cite{sidd-khe3}, the 
correlations of the shift and width values of the kaonic $^3$He and $^4$He $2p$ states
are plotted in Fig. \ref{results}, together with the average value reported in \cite{Baird, Batty-NPA},
where the error bars were calculated by adding the statistical and systematic errors  quadratically.
Clearly, the new results gave significantly smaller values both for the shift and width.

Presently, an upgrade of SIDDHARTA, SIDDHARTA-2,  is under way in order to 
perform the kaonic deuterium measurement. In the framework of SIDDHARTA-2 
we plan as well to challenge the difficult measurements of kaonic $^3$He and $^4$He 
transitions to the $1s$ level.

\section*{Acknowledgments}
\label{sec:ack}

We thank C. Capoccia, G. Corradi,
B. Dulach, and D. Tagnani from LNF-INFN; and
H. Schneider, L. Stohwasser, and D. St\"{u}ckler
from Stefan-Meyer-Institut,
for their fundamental contribution in designing and building the
SIDDHARTA setup.
We thank as well the DA$\Phi$NE staff for the excellent working
conditions and permanent support.
Part of this work was supported by
HadronPhysics I3 FP6 European Community program,
Contract No. RII3-CT-2004-506078;
the European Community-Research Infrastructure Integrating
Activity ``Study of Strongly Interacting Matter''
(HadronPhysics 2, Grant Agreement No. 227431), and
HadronPhysics 3 (HP3), Contract No. 283286
under the Seventh Framework Programme of EU;
Austrian Federal Ministry of Science
and Research BMBWK 650962/0001 VI/2/2009;
Romanian National Authority for Scientific Research,
Contract No. 2-CeX 06-11-11/2006;
the Grant-in-Aid for Specially Promoted Research (20002003), MEXT, Japan;
the Austrian Science Fund (FWF): [P20651-N20]; and
the DFG Excellence Cluster Universe of the Technische Universit\"{a}t M\"{u}nchen.

\bibliographystyle{elsarticle-num}
\bibliography{<your-bib-database>}

\end{document}